\documentclass[doublecol]{epl2}
\usepackage{amssymb}
\usepackage{amsmath}
\usepackage{graphicx}
\usepackage{color}
\usepackage{verbatim}

\newcommand{\beq}{\begin{equation}}
\newcommand{\eeq}{\end{equation}}
\newcommand{\beqs} {\begin{displaymath}}
\newcommand{\eeqs} {\end{displaymath}}
\newcommand{\beqa} {\begin{eqnarray}}
\newcommand{\eeqa} {\end{eqnarray}}
\newcommand{\beqas} {\begin{eqnarray*}}
\newcommand{\eeqas} {\end{eqnarray*}}

\newcommand{\Tensor}[1] {\mathbb{#1}}

\newcommand{\TensorT} {\mathcal{T\!\!\!\!T}}

\newcommand{\mean}[1]{\left\langle #1\right\rangle}

 \newcommand{\Label}[1]{\label{#1}}

\definecolor{magenta}{rgb}{0.7,0,0.7}

\title{Stochastic Quantization and Casimir Forces}

\author{P.~Rodriguez-Lopez\inst{1} \and  R.~Brito\inst{1} \and R.~Soto\inst{2}}
\shortauthor{P.~Rodriguez-Lopez \and R.~Brito \and R.~Soto}
\institute{\inst{1} Departamento  de F\'{\i}sica Aplicada I and GISC,
Universidad Complutense, 28040 Madrid, Spain\\
\inst{2} Departamento de F\'{\i}sica, FCFM, Universidad de Chile,
Casilla 487-3, Santiago, Chile}


\abstract{
In this paper we show how the stochastic quantization method
developed by Parisi and Wu can be used to obtain Casimir
forces. Both quantum and thermal fluctuations are taken into account by
a Langevin equation for the field. The method allows the Casimir
force to be obtained directly, derived from the stress tensor instead of
the free energy.
It only requires the spectral decomposition of the Laplacian operator
in the given geometry. The formalism provides also an expression for the fluctuations of the force.
As an application we compute the Casimir force on the plates of a
finite piston of arbitrary cross section. Fluctuations of the force are
also directly obtained, and it is shown that, in the piston case,
the variance of the force is twice the force squared.
}

\begin{document}
\maketitle

Fluctuation-induced (Casimir) forces~\cite{Casimir_Placas_Paralelas, Review_Casimir} are currently attracting renewed attention, probably because of the
accessability of small systems at the nano- and microscale, as it is
in these small systems where Casimir forces are revealed, despite some macroscopic effects~\cite{review2}.
However, techniques to calculate forces for complicated geometries,
beyond the usual ones with high degrees of symmetry, are scarce.
Recently, a powerful technique to calculate electromagnetic (EM)
Casimir forces~\cite{Kardar} has been proposed. It is based
on a multiscattering technique and has been successfully applied to
many configurations, such as plates, cylinders, spheres, wedges, etc.~\cite{Casimir_Wedges}

Casimir forces have their origin in fluctuations of EM fields.
Already Lifshitz in 1956 \cite{Lifshitz} use the concept of
fluctuations (in this case of the electric currents inside a metal) to 
compute such forces. 
With the  idea of fluctuations  in mind, some authors \cite{Ajdari,Fournier} used a method
that takes as its starting point the Langevin equation that describes the 
evolution of systems subjected to fluctuations (in their case  of thermal origin). 
Such method was applied to equilibrium situations, 
but also to non-equilibrium ones \cite{BartoloPRE,Najafi,We-PRE}, 
in which the Fluctuation-Disspiation theorem
is not satisfied.
However,  application of the Langevin approach to EM Casimir forces of quantum origin
has not yet been developed.

The goal of this paper is to apply the Langevin equation method to calculate forces
of quantum origin. The method is based on the stochastic quantization approach 
developed by 
 Parisi and Wu~\cite{Parisi-Wu_original}. These authors  construct a Langevin equation
for a given field subjected to thermal fluctuations, in a non-physical, pseudo-time.
Despite the classical nature of fluctuations we will show
how quantum fluctuations arise naturally within that method. 
We will follow the formalism developed in Ref.~\cite{We-PRE},
that takes as a starting point the Langevin equation derived by Parisi and Wu, 
and construct its solution by diagonalization of the deterministic evolution operator
with the appropriate bounday conditions ---the immersed bodies---, 
and later averaging over the thermal noise. 

The resulting method has the advantage of computing the Casimir force
directly and can be directly applied to compute torques and stresses
over extended bodies. Also, it is suitable for numerical or
perturbation methods, allowing to compute forces on realistic
geometries. Finally, the effects of finite temperature are included
directly.

The Casimir force is calculated via the stress tensor, $\Tensor{T}$ (see Ref.~\cite{We-PRE}), 
which must be averaged over the probability distribution
of the fields~$\phi$. Such a probability distribution is given by:
\beq\label{Pphi}P[\phi]= Z^{-1}e^{-S[\phi]/\hbar},
\eeq
where $S[\phi]$ is the action of the scalar field with zero mass,
Wick-rotated in the time variable~\cite{Review_Casimir} $(t = i\tau)$, i.e.,
\beq
S[\phi]= -\frac{1}{2} \int_0^{\beta\hbar} d\tau \int d{\bf r} \,
\phi \left(\frac{1}{c^2}\frac{\partial^2}{\partial \tau^2}+\nabla^2\right)\phi,
\eeq
where $\beta=1/k_BT$ and $Z$ is the partition function
$ Z= \int D\phi\, e^{-S[\phi]/\hbar} .$
For the bosonic case, the field~$\phi$ must obey periodic boundary conditions in time, that is,
$\phi(\tau+\beta\hbar,{\bf r})=\phi(\tau,{\bf r})$.
The stress tensor is normally a bilinear form in the field $\Tensor{T}=\TensorT[\phi,\phi,{\bf r}]$, the expression that defines the stress tensor operator $\TensorT$.

Herein we use the formalism of Parisi and Wu~\cite{Parisi-Wu_original, Masujima} to evaluate
the average of the stress tensor in an alternative way.
The idea of Parisi and Wu consists in a formulation of quantum mechanics or quantum field theory in terms of a stochastic process.
More precisely, a Langevin equation for the field~$\phi$ is written
as an evolution equation in an auxiliary time, which we will call pseudo-time, $s$.
In this description, the field depends on the new pseudo-time variable:
$\phi(\tau,{\bf r})\to \phi(\tau,{\bf r};s)$.
The Langevin equation takes the form
\begin{eqnarray}\label{Langevin}
\partial_s\phi (\tau ,{\bf r};s) & =& - \frac{\delta S[\phi]}{\delta \phi} +\eta(\tau ,{\bf r};s)\nonumber \\
 &= & \left(\frac{1}{c^2}\frac{\partial^2}{\partial \tau^2}+\nabla^2\right) \phi +\eta(\tau ,{\bf r};s).
\end{eqnarray}
The term $\eta(\tau ,{\bf r};s)$ is the source of fluctuations, given by
a~Gaussian white noise satisfying the fluctuation--dissipation relation~\cite{Kubo}
\begin{eqnarray}\Label{eq.noise}
\langle \eta (\tau ,{\bf r};s)\rangle& =& 0, \\
\langle \eta (\tau ,{\bf r};s)\eta (\tau' ,{\bf r}';s')\rangle& =&
2k_BT \delta(\tau-\tau')\delta ({\bf r}-{\bf r}')\delta(s-s'). \nonumber
\end{eqnarray}
Higher correlation of odd number of $\eta$-functions vanish, while correlations of an even
number of $\eta$-functions factorize in all possible products of two $\eta$-functions \cite{Risken}. For instance, the 4-correlation:
\begin{eqnarray}\Label{correlations}
\langle \eta (1)\eta (2)\eta(3)\eta(4)\rangle =
\left(2 k_B T\right)^{2} 
\big[\delta(1-2)\delta(3-4)&&\nonumber \\+\delta(1-3)\delta(2-4)+\delta(1-4)\delta(2-3)\big],&&
\end{eqnarray}
where the symbol 1, 2, 3 or 4 stands for the complete set $\tau, {\bf r}, s$, and the $\delta$ means the product of 3 $\delta$-functions, as in
Eq. (\ref{eq.noise}).

The solution of the Langevin equation in the stationary limit $s\to \infty$ reproduces the probability distribution
given by Eq.~(\ref{Pphi})~\cite{Masujima}. The pseudo-time $s$ has not physical meaning and the role of the evolution in $s$ in solely to sample over the probability distribution (\ref{Pphi}). Statistical correlations of two or more fields, must be evaluated at the same value of $s$, but otherwise the positions and times $\tau$ can be arbitrary.

This approach, although similar, is qualitatively different to the Lifshitz method. In Lifshitz method, fluctuating electric currents are introduced in the conductors which produce the fluctuating electromagnetic field in the vacuum. The fluctuations of the current are thermal, taking place in real time, and the quantum character of the model is introduced by the use of the quantum fluctuation-dissipation relation \cite{Lifshitz}. In the Parisi-Wu method, the fluctuation process takes place in a pseudo-time and its role is to reproduce the full quantum-thermal statistical properties of the field. In particular, the complete quantum correlations, including quantum coherence, are obtained. The second difference is that the Parisi-Wu approach does not consider the fluctuations in the macroscopic metal but on the (electromagnetic) field itself.

Having an expression for the stress tensor and a Langevin equation for the field, we can follow the procedure
recently developed in~\cite{We-PRE} to obtain the Casimir force. The field~$\phi$ is written as
(and a similar decomposition
for the noise~$\eta$ with coefficients $\eta_{nm}$):
\beq
\phi (\tau ,{\bf r};s)=\sum_{n,m} \phi_{nm}(s) g_m(\tau) f_n({\bf r}), \Label{eq.phi}
\eeq
where $f_n$ and $g_m$ are the eigenfunctions:
\beq
\nabla^2 f_n({\bf r})=-\lambda_n^2 f_n({\bf r}), \,\,\, \frac{1}{c^2}\frac{\partial^2}{\partial \tau^2}g_m(\tau) =-\omega^2_m g_m(\tau),
\label{eq.eigenfunctions}
\eeq
which are orthogonal under the L$^2$ scalar product in space
or time. The above expression indicates that $f_n({\bf r}) $ and $\lambda_n^2$
encode the spatial configuration of the system, that is, the position of the bodies and their boundary conditions.
In a similar fashion, $g_m(\tau)$ and $\omega^2_m$ contain the (Wick-rotated) time dependence.
As we are considering a bosonic field that obeys periodic boundary conditions in~$\tau$,
the eigenvalues are the known Matsubara frequencies $\omega_m=2\pi m/\beta\hbar c$, $m\in\mathbb{Z}$
, and the eigenfunctions are
$g_{m}(\tau)=\exp(-i\omega_{m}\tau)$. Then, from Eq. (\ref{Langevin}), the coefficients $\phi_{nm}$ satisfy the
differential equation
\beq
\frac{d\phi_{nm}(s)}{ds}=-  \left[\lambda_n^2+\omega_m^2\right]\phi_{nm}(s)+ \eta_{nm}(s),
\eeq
which can be integrated to give
\beq
\phi_{nm}(s)=\int_{-\infty}^s d{\sigma} e^{(\lambda_n^2 +\omega_m^2)(\sigma-s)} \eta_{nm}(\sigma), \label{phieta}
\eeq
where the noise coefficients satisfy, from Eq. (\ref{eq.noise})
\beq
\langle \eta_{nm}(s)\eta^*_{n'm'}(s')\rangle= 
2k_BT \delta(s-s') \delta_{nn'}\delta_{mm'}. \label{corr.eta}
\eeq
The formalism developed in~\cite{We-PRE} allows the average stress tensor to be calculated by substituting expression
(\ref{eq.phi}) into the stress tensor and taking the average
over the fluctuations, $\eta(\tau,{\bf r};s)$,
in the limit $s\to\infty$:
\begin{eqnarray}
\langle\Tensor{T}({\bf r})\rangle&=& 
\lim_{s\to\infty}\langle\TensorT[\phi(s),\phi(s),{\bf r}]\rangle \\
&=& 
\lim_{s\to\infty}
\sum_{\begin{minipage}{0.8cm}${\scriptstyle{n_1,m_1}}$\\[-2mm]${\scriptstyle n_2,m_2}$\end{minipage}} 
\langle \phi_{n_1m_1}(s)\phi_{n_2,m_2}^*(s)\rangle
\TensorT[f_{n_1},f_{n_2}^*,{\bf r}]. \nonumber 
\end{eqnarray}
The two functions $\phi_{nm}$ are replaced by their values in terms of the 
fluctuations, given by Eq. (\ref{phieta}), resulting into a double integral over two pseudo-times.
Then, the average $\langle\cdot\rangle$ is carried out using Eq. (\ref{corr.eta}), that eliminates
one integral, with the final result:

\begin{equation}
\langle\Tensor{T}({\bf r})\rangle =
\frac{1}{\beta}\sum_{nm}\frac{\TensorT_{nn}({\bf r})}
{\lambda_n^2+\omega_m^2}, \label{eq.T}
\end{equation}
where $\TensorT_{nm}({\bf r})=\TensorT[f_n,f_m^*,{\bf r}]$. 
%
Moreover,    the sum over the Matsubara frequencies, $\omega_m$, can
be carried out to obtain the result 
\beq 
\langle \Tensor{T}({\bf
r})\rangle= \frac{\hbar c}{2}\sum_{n} \frac{\TensorT_{nn}({\bf r})}
{\lambda_n} \left[1+\frac{2}{e^{\beta \hbar c \lambda_n}-1}\right],
\Label{eq.T2}
\eeq which can be related with the quantum
fluctuation--dissipation theorem for the EM field~\cite{Lifshitz,QFDT_Weber}.
Finally, to obtain the Casimir force over a~certain body, the stress
tensor must be integrated over the surface~$\Omega$ that defines the
object
\begin{equation}  \Label{eq.FCint}
{\bf F}_{C}  = \oint_\Omega \langle \Tensor{T}({\bf r})\rangle\cdot d{\bf S} . 
\end{equation}

Equation~(\ref{eq.T2}) is the main result of this letter. It
gives an expression for the quantum Casimir force including the
effects of a finite, nonvanishing temperature, in terms of the eigenvalues and eigenfunctions of the
Laplacian operator. This expression has been easily obtained using the
stochastic quantization approach to a quantum field, together with the formalism of the Langevin equation
calculation of fluctuation-induced forces~\cite{We-PRE}. 

Let us note that, as it occurs in most of the calculations of the Casimir Force, 
the expression for $\langle \Tensor{T}({\bf r})\rangle$ in Eq. (\ref{eq.T2}) is generally 
divergent at every point of space ${\bf r}$ if the sum over eigenvalues runs to infinity. 
However, the expression for the force, that is obtained integrating over the surface of 
the body, Eq.(\ref{eq.FCint}), is finite. It means that the integration over the body 
regularizes the divergences of the averaged stress tensor. If we assume that 
such regularization is carried out mode by mode, we can interchange the integration over the surface 
and the summation over eigenvalues, to obtain,
\beq
{\bf F}_{C} = \frac{\hbar c}{2}\sum_{n} \frac{1}
{\lambda_n} \left[1+\frac{2}{e^{\beta \hbar c \lambda_n}-1}\right]
\oint_\Omega \TensorT_{nn}({\bf r}) \cdot d{\bf S}.
\Label{eq.T3}
\eeq 
which is a {\em finite} result. Therefore, the interchange of the integral and summation
regularizes the Casimir force, avoiding the use of ultraviolet cutoffs. 
Other regularizations, that in some cases may lead to non-universal forces or fluctuations
are, for instance, the subtraction of the vacuum stress tensor \cite{Ford} or
by averaging the stress tensor over a finite area or a finite time \cite{Barton}.
Having regularized the divergences, Eq. (\ref{eq.T3}),
provides a new method to calculate Casimir forces for a~given geometry by diagonalizing the
Laplace operator. So, this approach is suitable for numerical calculations of Casimir forces
in complicated, realistic geometries. Moreover, this method provides the force directly, not
as a difference of the free energy with respect to a reference state, which in some configurations it may be
difficult to establish. Also, it can be used as a starting point for a
perturbative theory for, e.g., nonflat geometries, rough surfaces or similar problems. 
Other authors 
have obtained expressions for the free energy in terms of the 
eigenvalues of the spatial operator, but not for the force~\cite{Marachevsky,Abalo}.

The expression for the Casimir force, Eq.~(\ref{eq.T2}), allows one
to evaluate the quantum limit, that is, by setting the temperature
equal to zero (or $\beta\to \infty$), where the second summand
inside the brackets in Eq.~(\ref{eq.T2}) vanishes, i.e., \beq
\Label{eq.FCT0} \lim_{T\to 0} \langle \Tensor{T}({\bf r})\rangle=
\frac{\hbar c}{2}\sum_{n} \frac{\TensorT_{nn}({\bf r})} {\lambda_n}.
\eeq In the opposite, classical limit, when $\hbar \to 0$, a Taylor
expansion of the square bracket in Eq.~(\ref{eq.T2}) gives
\beq\Label{eq.FCh0} \lim_{\hbar \to 0} \langle \Tensor{T}({\bf
r})\rangle= \frac{1}{\beta} \sum_{n}\frac{\TensorT_{nn}({\bf r})}
{\lambda_n^2}. \eeq This expression for the Casimir force has been
used for classical systems~\cite{We-PRE}, such as liquid
crystals~\cite{Ajdari} or reaction--diffusion
systems~\cite{BritoSotoPRE}. The two limits, quantum (\ref{eq.FCT0})
and thermal (\ref{eq.FCh0}), show that the driving force of the
fluctuations has different origin. In the first case, the presence
of the factor~$\hbar$ indicates the quantum nature of the
fluctuations, whereas in the second case, the factor $\beta$ reveals
its thermal origin.

{\em Fluctuations}. The Casimir force has its origin in
fluctuations, so it is a fluctuating quantity itself. The net force
is calculated as the average of the fluctuating one,
Eq.(\ref{eq.FCint}). However, the formalism developed in this letter
allows the calculation of fluctuations of the force, defined as
customary for the fluctuations: \beq\Label{Fluc} \sigma_F^2=
\oint_\Omega  \oint_\Omega \langle[\Tensor{T}({\bf r}_1)\cdot d{\bf
S}_1] [\Tensor{T}({\bf r}_2)\cdot d{\bf S}_2]\rangle -F_C^2. \eeq
Calculation of $\sigma_F^2$ requires the evaluation of the product
of 4 fields $\phi$, as the tensor $\Tensor{T}$ is bilinear form of
$\phi$. Substitution of $\phi$ in the tensor leads to the
correlation of a product of four noises, $\eta$. Because of the
Gaussian nature of $\eta$, it factorizes into three products of
pairs of noises, as shown in Eq.~(\ref{correlations}). The precise
expression for the product of the two stress tensors in Eq
(\ref{Fluc}) reads:
\begin{eqnarray}\Label{integrand}
\langle\Tensor{T}({\bf r}_1)\Tensor{T}({\bf
r}_2)\rangle=\frac{(\hbar c)^2}{4}\sum_{nm}P(\lambda_n)P(\lambda_m)\times\hspace*{1.5cm}&&\nonumber \\
 \hspace*{1cm}\left[ \TensorT_{nn}({\bf r_1})\TensorT_{mm}({\bf r_2})
+2\TensorT_{nm}({\bf r_1})\TensorT_{mn}({\bf r_2}) \right], &&
\end{eqnarray}
where
 \beq P(\lambda)= \frac{1}{\lambda}\left[1+\frac{2}{e^{\beta
\hbar c \lambda}-1}\right]. \eeq
The first term in (\ref{integrand})
exactly gives $F_C^2$ when substitued in Eq.(\ref{Fluc}), while the other term (where the factor 2
comes because the two last terms in Eq.(\ref{correlations}) give
equal terms) give nonvanishing contributions to $\sigma_F^2$.
Again, there is no need to introduce cutoff in the
eigenvalues, as interchanging the summations with the integral over the bodies
regularize the divergences. Therefore, we find a universal form for the fluctuations, 
as opposite to other authors \cite{Fournier,Barton}. 
The difference has its origin in that we do not compute the stress
fluctuations or the fluctuations of the force on each side of a plate.
Rather, we first compute the total fluctuating force on the body
(which is finite) and then we compute its variance.

\begin{figure}[htb]
\includegraphics[width=0.98\columnwidth,angle=0]{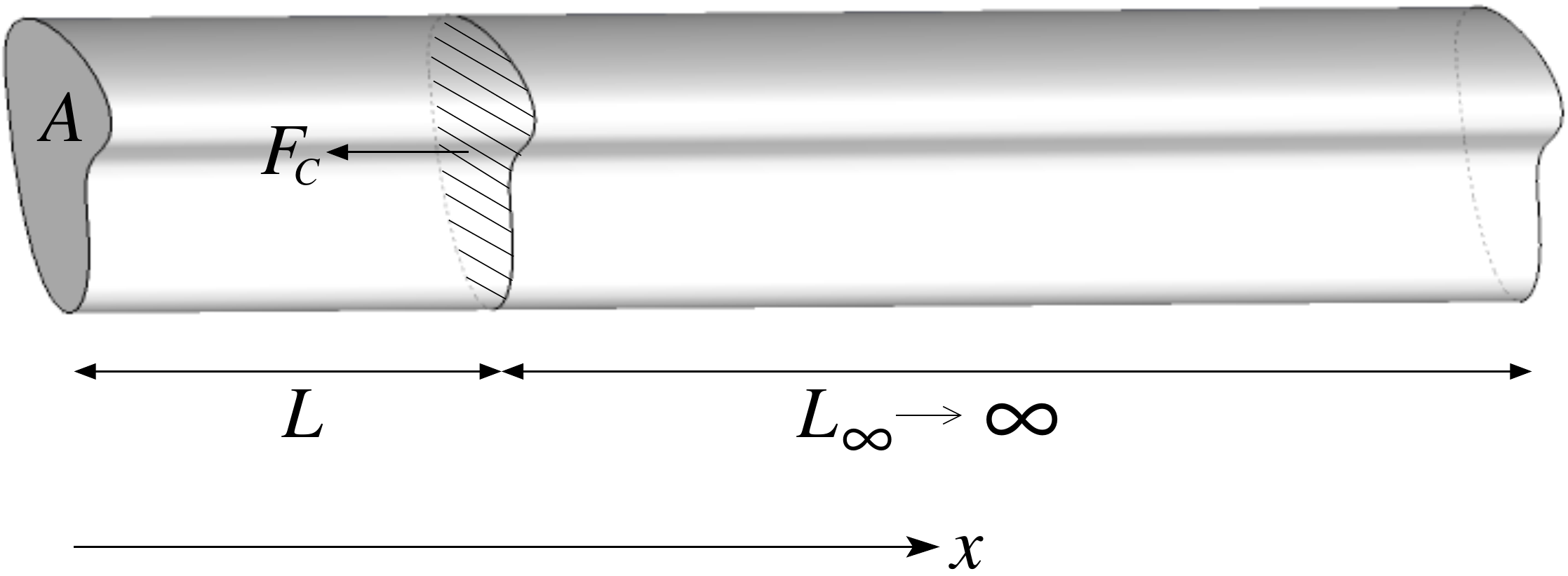}
\caption{Geometry of the considered problem. The Casimir force is evaluated for the plate, of area $A$ and general geometry.
The net force is obtained as the force exerted by a plate at distance $L$ and another plate at distance $L_\infty\to\infty$.
The cylinder is oriented along the $x$ axis.}
\label{fig.plates2}
\end{figure}

{\em Application to the calculation of the Casimir force of a  cylinder of arbitrary cross section}.
In order to show how the formalism developed in this letter works, let us consider a simple case of a piston depicted in
Fig. \ref{fig.plates2}. It consists of a piston of area $A$, but general shape,
made of a perfectly conducting metal surface \cite{Hertzberg}.
Two flat conducting plates of the same cross section of the piston are placed at a~distance~$L$ apart along the $x$ direction.
The plates are perpendicular to the surface of the cylinder.
We calculate the Casimir force, $F_C$, for this plate by evaluating  the expressions Eq. (\ref{eq.T2}) and later integrating over the surface.
In order to obtain a finite result, we need
a third auxiliary plate located at infinite distance, $L_\infty\to\infty$.

First, we solve the eigenvalue problem for the EM field and apply
Eqs.~\eqref{eq.T} and \eqref{eq.FCint} to obtain the force.
For the EM field the normal component of the stress tensor  reads
$ \mathbb{T}_{xx}=E_{x}^{2} + B_{x}^{2} - \frac{1}{2}\textbf{E}^{2}- \frac{1}{2}\textbf{B}^{2},
$
which has to be averaged over the noise and then integrated over the surface of the plates, as shown in Eq.~\eqref{eq.FCint}.
For perfectly conducting plates, the boundary conditions are:
$\textbf{E}\times\textbf{n} = \textbf{0}$ and $\textbf{B}\cdot \textbf{n}=0$, where $\textbf{n}$ is the normal
vector at the surface.
In this geometry, the EM field can be decomposed into transverse electric (TE)
and transverse magnetic (TM) modes, which are discussed independently~\cite{Jackson}.

For the TM modes, the magnetic field is transversal to the $x$ direction, and
the vector potential ${\bf A}$ for TM modes can be written as
\beq
{\bf A}=( -C \nabla_\perp^2 D, \partial_y D\, \partial_x C, \partial_zD\,\partial_x C ) e^{-i\omega t}.
\eeq
Here the fields $C(x)$ and $D({\bf r}_\perp)$ satisfy
\begin{eqnarray}\Label{eq.Cg}
\partial_x^2C(x)=&-k_x^2 C(x), &{\text{\ (Neumann BC on\ }} x=0,L{\text{)}} \nonumber  \\
\nabla_{\perp}^{2}D_{n} ({\bf r}_\perp)=& - \lambda_{n}^{2}D_{n}({\bf r}_\perp),  &{\text{\ (Dirichlet BC on  }} {\cal S} {\text{)}},
\end{eqnarray}
where ${\cal S}$ is the surface of the cylinder and ${\bf r}_\perp=(y,z)$.
For the TE set, the electric field is transversal to $x$, so the vector potential is
$ {\bf A}=( 0, -S\,\partial_z N, S\, \partial_yN  ) e^{-i\omega t} $,
where the functions $S(x)$ and $N({\bf r}_\perp)$ satisfy Eqs.~(\ref{eq.Cg})
with the opposite boundary conditions: Dirichlet for $S(x)$ and Neumann for $N({\bf r}_\perp)$.
However, in this case, we must exclude the constant eigenfunction, with eigenvalue $\lambda_n^2$ equal to zero, as
it gives that ${\bf A}={\bf 0}$ and then ${\bf E}={\bf B}={\bf 0}$.

Substitution of the TE modes into the expression for the stress tensor, and integration over one side of the
plates gives, after a long but straightforward calculation, the Casimir force  as
\begin{equation}\Label{12}
\int_{\tiny{\text{1 side}}}\hspace{-0.4cm}\mean{\mathbb{T}_{xx}^{TE}}
dS_x =
\frac{2}{\beta L}\sum_{m\in\mathbb{Z}}\sum_{n_{x}=1}^{\infty}\sum_{n}
\frac{k_{x}^{2}}{\,\omega_{m}^{2} + k_{x}^{2} + \lambda_{n}^{2}\,},
\end{equation}
where $k_{x}^{2}= (n_x\pi/L)^2$, and $\omega_m$ is defined after Eq.~\eqref{eq.eigenfunctions}.
For the TM modes, one obtains exactly the same expression, but $\lambda_{n}^{2}$ are the eigenvalues of the two-dimensional (2D) Laplacian
with Neumann boundary conditions. We will denote the complete set of eigenvalues of
the Laplacian with Neumann (excluding the zero eigenvalue)
and Dirichlet boundary conditions by the index~$p$.
The expression above is the equivalent of Eq.~\eqref{eq.T} when the spectrum can be split into a~longitudinal and transversal part, that is, $\lambda_n^2= k_x^2+\lambda_p^2$.
These series are divergent, but the net Casimir force,
which is the difference between the force exerted from the plate at distance $L$, and
the plate at $L_\infty\to\infty$ (see Fig. 1), is finite.

The sum over the variable $n_x$ in Eq.~(\ref{12}) can be carried out with the help of the
Chowla--Selberg summation formula~\cite{Elizalde}. This formula extracts the divergent, $L$-independent
part of the summation, which cancels when the integral in Eq.~(\ref{12}) is
performed for both sides of the piston. Equivalently, the $L$-independent contribution is the same for
the second plate at distance $L$ or at distance $L_\infty$,  resulting in
\beq\Label{FT}
F_{C} = - \frac{1}{\beta}\sum_{p}\sum_{m\in \mathbb{Z}} \frac{\sqrt{m^2 \Lambda^2+\lambda_p^2}}{e^{2 L \sqrt{m^2 \Lambda^2+\lambda_p^2}} - 1}.
\eeq
Here, $\Lambda=2\pi /\beta\hbar c $ is the inverse thermal wavelength. This expression gives the finite or regularized Casimir force between two plates
at distance~$L$, valid for any cross section and temperature. The precise geometry of the plates enters into the double
set of eigenvalues of the Laplacian (with Neumann and Dirichlet boundary conditions)
$\lambda_p^2$.

We can proceed to evaluate the Casimir force at $T=0$, that is, the purely EM case
without thermal corrections.
This case is obtained by noting that, when $T\to 0$ [equivalent to the limit $\Lambda\to 0$ in Eq.~(\ref{FT})], the sum over~$m$ can be rigorously replaced by an integral. Computing the integral for finite~$\Lambda$ and taking the limit $\Lambda\to 0$, the result is
\beq\Label{eq.FCBessel}
 F_C= - \frac{\hbar c}{2\pi}
 \sum_{p}\sum_{n=1}^\infty \lambda_p^2   \left[ K_0(2n L \lambda_p )+
  K_2(2n L \lambda_p)\right] .
 \eeq
Here $K_{\alpha}(x)$ is the modified Bessel function of order~$\alpha$.
In Ref.~\cite{Marachevsky} it was obtained a formula for the free energy of the configuration considered here, that, after differentiation with respect to the distance between the plates, leads to 
the force above.

For short distances, however, the summation above can be calculated
without explicitly knowing the eigenvalues of the Laplacian. Such eigenvalues,
$\lambda_p^2$, 
must scale with the inverse of the square of the typical size of the piston. So, for
distances~$L$ much smaller than the section of the piston, we can
replace the sum over the eigenvalues, $\lambda_p^2$, by an integral, using the
asymptotic expression for the density of states of the
 Laplacian in two dimensions: $\rho(\lambda_p) = \frac{A}{2\pi}\lambda_p,$~\cite{DoS}  for each set of Dirichlet or Neumann.
 This result will be universal, independent of the shape of the section of the piston, as the density of states is itself universal.
The resulting integrals can be performed to obtain the Casimir force as
\begin{equation}\Label{eq.Near}
F_{T=0} = - \frac{\hbar c\pi^{2}}{240L^{4}}A,
\end{equation}
which is the well-known result of the EM Casimir force for infinite  parallel plates~\cite{Casimir_Placas_Paralelas}.
In the opposite limit, when~$L$ is much larger than the typical size of
the plate, the argument of the Bessel functions in Eq.~(\ref{eq.FCBessel}) is much larger than one.
Because of the exponential behavior of the Bessel functions,
only the smallest eigenvalue~$\lambda_{1}^2$ contributes to the sum, with the result
\begin{equation}\Label{eq.Far}
F_{T=0} = - \frac{\hbar c}{2\sqrt{\pi L}}g_{1}
\lambda_{1}^{3/2}e^{-2L\lambda_{1}},
\end{equation}
being $g_{1}$ the degeneration of $\lambda_{1}$.
Here, a counterintuitive result is obtained.
One would expect that the thin piston would tend to the known one-dimensional (1D) Casimir force,
but instead an exponential decay of the force is found. The known 1D result would be obtained if the zero eigenvalue
were considered. However, this eigenvalue is excluded, as it
leads to a vanishing eigenfunction.
Therefore, the 1D Casimir force cannot be obtained as the limit from three dimensions (3D) to 1D.

At intermediate distances, that is, $L$ comparable to the size of the plates, one must
solve the eigenvalue problem. To illustrate the intermediate behavior and the transition from
Eq.~(\ref{eq.Near}) to (\ref{eq.Far}), we study the case of a circular cylinder of radius~$R$.
In this case, the eigenvalues are the zeros of the Bessel function $J_\nu(r)$ 
and its derivative.
The dependence with the radius of the cylinder is 
$\lambda_p^2(R)=\lambda_p^2(R=1)/R^2$, so the Casimir force as expressed in Eq. (\ref{eq.FCBessel})
is a function of $L/R$ when the force is multiplied by $R^2$.

\begin{figure}[htb]
\includegraphics[width=0.98\columnwidth,angle=0]{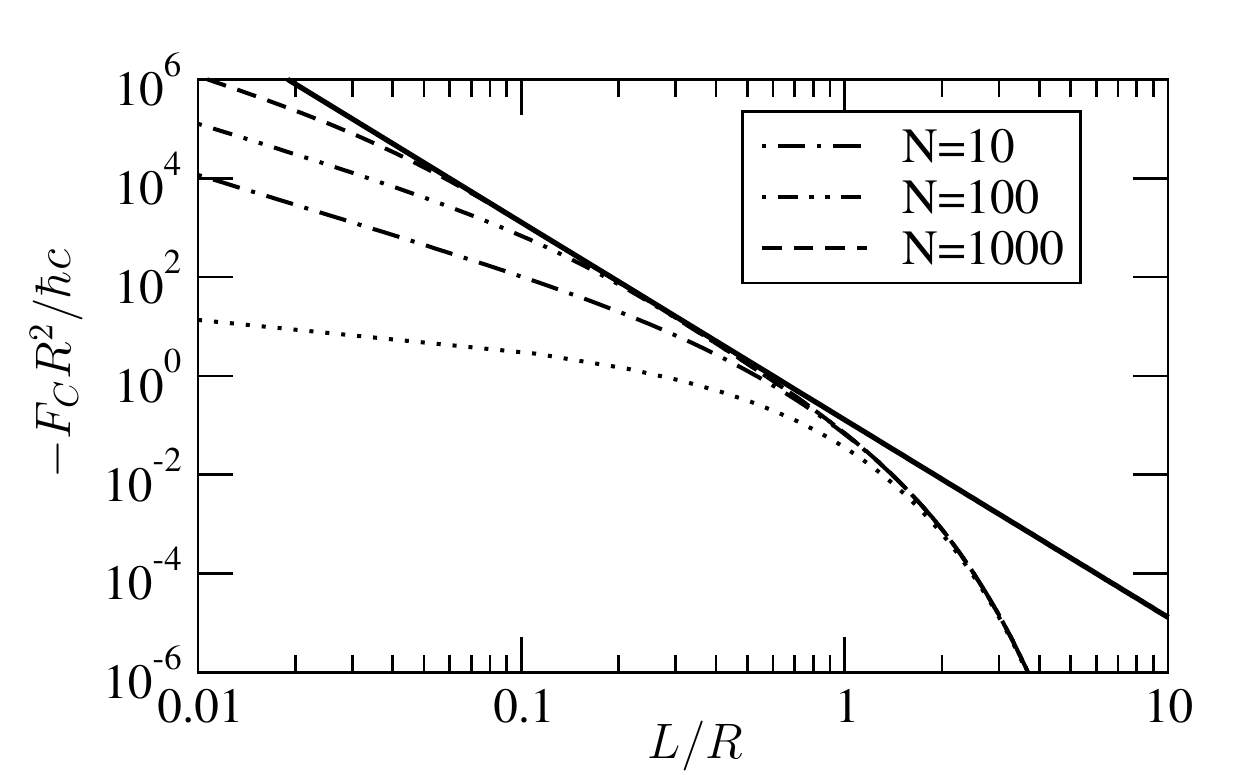}
\caption{Casimir force on the plates of a~cylindrical piston of circular cross section of radius~$R$ as a function of the distance~$L$ between the plates.
Different curves are obtained by summing $N$
eigenvalues of the Laplacian, using Eq.~(\protect{\ref{eq.FCBessel}}), where $N$ is
indicated in the legend. Solid  line is the 3D Casimir force, Eq.\protect{(\ref{eq.Near})}
and dotted  line is the far distance result Eq.\protect{(\ref{eq.Far})}.
}
\label{fig.plates}
\end{figure}

Figure \ref{fig.plates} shows  the results  for the Casimir force, Eq. (\ref{eq.FCBessel}),
 when $N=10, N=100$ and $N=1,\!000$ eigenvalues are summed (taking  into account their
degeneration). We have also plotted
the two limiting results: (i) the 3D Casimir force (\ref{eq.Near}),
valid for $L\ll R$ with an algebraic behavior: $F_C R^2\propto (L/R)^{-4}$ (solid  line), (ii) 
the far distance limit, given by (\ref{eq.Far}) with the smallest eigenvalue
$\lambda_1^2(R)\simeq 3.39/R^2$ with degeneracy $g_1=2$,
that is valid for $L\gg R$ (dotted   line).
The transition between both regimes is observed at  $L\simeq R$.
As expected, when few eigenvalues are summed, for instance $N=10$, 
the resulting force is only valid in the limit of long distances. As the number of
eigenvalues increase, the numerical result approaches the 3D Casimir force. 
For $N=1,000$ eigenvalues, we have excellent results for $L/R>0.05$.
We remark, however, that the full curve can be obtained by only considering $N=100$ 
eigenvalues for large distances and matching this numerical result with the  
asymptotic expression (\ref{eq.Near}) for short distances, 
with a crossover distance $L\approx0.3 R$.

In a similar fashion, we can calculate the thermal Casimir force when $\hbar \to 0$,
or $\Lambda\to\infty$. Then, in Eq. (\ref{FT}), only the term with $m=0$
is different from zero, as a result,
\begin{equation}\Label{eq.ant}
F_{\hbar=0} = - \frac{1}{\beta}\sum_{p}\frac{\lambda_{p}}{e^{2L\lambda_{p}} - 1}.
\end{equation}
For short ($L\!\ll\! R$) and long ($L\!\gg\! R$) distances,
(\ref{eq.ant})  reads
\begin{equation}
F_{\hbar=0} = - \frac{\zeta(3)}{4\beta \pi L^{3}}A ,\quad
F_{\hbar=0} = - \frac{1}{\beta}g_{1}\lambda_{1}e^{-2L\lambda_{1}}.
\end{equation}

We finish the application to the cylinder by evaluating the fluctuation of the Casimir Force.
For the piston geometry considered in this letter, the fluctuations of the force can be obtained by evaluating
Eq.~(\ref{Fluc}). In this case, and because of the geometry of the problem, each
of the summands that appear in the four-point correlations of the noise
gives~$F_C^2$. Therefore, the fluctuations of the force, for any temperature and cross section, are simply
\beq
\sigma_F^2=2  F_C^2.
\eeq
Somehow similar fluctuations have been obtained for a~purely thermal force~\cite{Fournier} 
for each mode that enters in the Casimir force. 
In a non-equilibrium hydrodynamical system the fluctuations have been 
measured in~\cite{Cattuto} by means of numerical simulation.
The fact that the fluctuation of the force is as large as the force itself is
a signature of fluctuation-induced forces.

{\em Summary.} We have shown in this letter that stochastic quantization
together with the Langevin formalism provides a new approach to calculate Casimir
forces in the quantum electrodynamics (QED) case, including thermal effects. The starting point is the calculation
of the Casimir force via the stress tensor, which is a function of the
fluctuating fields. Parisi and Wu derived a Langevin equation to describe
the dynamics of such fields, which can be integrated to give an expression
for the force. The method presented herein
is quite simple, and avoids some technical complication of other approaches.
Moreover, it calculates the force directly, instead of the free energy.
It also can be extended to calculate torques in asymmetric configurations.
Another advantage is that it provides a numerical method to calculate
the Casimir force in complicated geometries, such as those of interest for microelectromechanical systems (MEMS)
devices, alternative to others \cite{Casimir_Numerico}. 
The method only requires spectral decomposition of the Laplacian
operator in the given geometry, and summation of the eigenvalues and the integration of the eigenfunction
along the boundary of the object, as shown in Eqs.~(\ref{eq.T2}) and (\ref{eq.FCint}).
Quantum ($T\to 0$)  and classical ($\hbar\to 0 $) limits are recovered by Eqs.~(\ref{eq.FCT0}) and
(\ref{eq.FCh0}) respectively. 
The integration of the eigenfunction over the surface of the body leads to a regularization of
the Casimir force, producing finite results for the force and the fluctuations.


This approach can be generalized to include non equilibrium situations. For instance, it can be used to calculate the Casimir force between two bodies at different temperatures \cite{Antezza,Emig-thermal}. Also, the method could be applied to time evolving temperatures~\cite{Dean-evolving_temperatures} and space dependent temperature $T({\bf r})$.

\acknowledgments{
The authors would like to thank
A.~Mu\~noz-Sudupe and A.~N\'u\~nez for helpful discussions.
P.R.-L. and R.B. are supported by the Spanish projects
MOSAICO and MODE\-LICO.
 P.R.-L.'s research is also supported by an FPU MEC grant.
R.S. is supported by Fondecyt grant 1100100
and Proyecto Anillo ACT 127. }



\begin{thebibliography}{100}
\bibitem{DoS} The asymptotic DoS has corrections coming from the perimeter of the piston and
its curvature \protect{\cite{Hertzberg}}. 
\end{thebibliography}
\end{document}